\documentclass[usegraphicx,usenatbib,useAMS]{mn2e}
\usepackage{times}
\usepackage{amssymb}
\usepackage{bm}

\newcommand{\be}{\begin{equation}}
\newcommand{\ee}{\end{equation}}
\newcommand{\ba}{\begin{eqnarray}}
\newcommand{\ea}{\end{eqnarray}}
\newcommand{\bc}{\begin{center}}
\newcommand{\ec}{\end{center}}

\newcommand{\nn}{\nonumber\\}

\newcommand{\bC}{{\bf C}}
\newcommand{\bF}{{\bf F}}

\newcommand{\bb}{{\bf b}}

\newcommand{\bn}{{\bf n}}

\newcommand{\bx}{{\bf x}}

\newcommand{\bbmu}{\mbox{\boldmath $\mu$}}

\newcommand\rstar{R_{\star}}

\def\gs{\mathrel{\raise1.16pt\hbox{$>$}\kern-7.0pt 
\lower3.06pt\hbox{{$\scriptstyle \sim$}}}}         
\def\ls{\mathrel{\raise1.16pt\hbox{$<$}\kern-7.0pt 
\lower3.06pt\hbox{{$\scriptstyle \sim$}}}}         

\begin{document}

\title[Fast transit identification]{Fast identification of transits from light-curves}
\author[Protopapas, Jimenez \& Alcock]
{Pavlos Protopapas$^1$, Raul Jimenez$^2$ and Charles Alcock$^1$\\
$^1$Harvard Smithsonian Center for Astrophysics, 60 Garden Street, Cambridge, MA 02138, USA.\\
$^2$Dept. of Physics and Astronomy, University of Pennsylvania, Philadelphia,
PA 19104, USA.}

\maketitle

\begin{abstract}
 We present an algorithm that allows fast and efficient detection of
 transits, including planetary transits, from light-curves. The method is based on
 building an ensemble of fiducial models and compressing the data
 using the MOPED algorithm.  We describe the method and demonstrate
 its efficiency by finding planet-like transits in simulated Pan-STARRS
 light-curves.  We show that that our method is independent of the size of the search space of transit 
parameters. In large sets of light-curves, we achieve speed up
factors of order of $10^{8}$ times over the full $\chi2$
 search.  We discuss how the algorithm can be used in forthcoming
 large surveys like Pan-STARRS and LSST and how it may be optimized 
for future space missions like Kepler and COROT where most of
 the processing must be done on board.
\end{abstract}


\section{Introduction}

If the orbit of a planet around a star is so favorably inclined that
$\sin(i) \approx 1$, the planet will transit the disk of the star once
per orbit. During the transit the observed flux from the star is
reduced by the ratio of the areas of the planet and the star,
typically $\sim 1$\% for a Jupiter-like planet around a Sun-like
star. When this photometric dimming is observed to repeat
periodically, a small radius companion may be inferred to exist.  This
effect is seen in star HD209458 \citep{Charbonneau2000}, which was
first identified as a planetary system using the radial velocity
technique.  The added value of the detection of transits is great: not
only is the $\sin(i)$ ambiguity resolved, but the radius of the planet
may be inferred, and spectroscopic examination of the object during
transit allows the study of the atmosphere of the planet
\citep{Charbonneau2002}.

The transit technique to search for planets has some advantages:
photometry is less costly in telescope time than spectroscopy, and one
knows $\sin(i)$ for all the systems found this way. The major
disadvantage is that the yield is comparatively low, since only
systems with $\sin(i) \approx 1$ will be detected.

A large number of transit searches for extra-solar planets, both
space-based and ground-based, have been completed or are underway
\citep{Gilliland.2000,Mochejska.2002,Udalski.2002,Udalski.2003,Mallen.2003}.
Many of these efforts employ small-aperture, wide-field cameras to
monitor tens of thousands of nearby, bright stars. Of the surveys
using this approach, the only success to date has come from the
Trans-atlantic Exoplanet Survey (TrES), which recently announced the
discovery of a planet dubbed TrES-1 \citep{Alonso2004}.  The only
other success (albeit for fainter stars where follow-up is more
difficult) has come from the OGLE survey
\citep{Udalski.2002,Udalski.2003}. The vast majority of transits have
been false detections resulting from grazing transits of stellar
companions or a blend of an eclipsing binary with a brighter
foreground or background star \citep{Torres.2004, Pont.2005}. Some progress has
been made at differentiating these from planets \citep{Hoekstra.2005}.
However, the few candidates not eliminated 
by follow up studies, in particular OGLE-TR-56b with its 1.2-day
orbital period, further challenge our already revised models of planet
formation \citep{Konacki2003}.

The detection of a weak, short, periodic transit in noisy light-curves
is a challenging task. The large number of light-curves collected make
automation and optimization processes a necessity. This requirement
is even stronger in the context of space missions where much of the
processing must be done on board.  A number of transit-detection
algorithms have been implemented in the literature
\citep{Doyle+00,Defay+01,AigrainFavata02,Jenkins+02,Kovacs+02,Udalski.2002,Street+03}
and there has been some effort to compare their perspective
performances \citep{Tingley2003} .  

Such transit searches are generally performed by comparing light-curves to
a family of models with a common set of parameters: the transit period $T$, the transit duration $\eta$, 
the epoch $\tau$ (which is equal to the time $t$ at the start of the first transit) 
and the transit depth $\theta$. 
The best set of parameters is identified by finding the model most likely to have
given rise to the observed data, i.e. the model with the highest likelihood
$L$. This is exactly the kind of problem MOPED \citep{HJL00} was
designed to address. In particular, light-curves contain plenty of
redundant information: the light between transits. By using MOPED one
can weigh more the part of the light-curve that is sensitive to the
transit thus constructing one eigenvector for each of the parameters
in the transit model.  However, for the case of transit detection in
light-curves, the MOPED eigenvectors are sensitive to the fiducial
model, thus incorrectly overweighs some data . In this paper we present a solution to this problem by building
an ensemble of fiducial models.  We find that for each model in an ensemble of
fiducial models, there are many possible solutions. 
However, only one solution is common to all  models in the ensemble
of fiducial models: the one with the correct parameter values of the
transit.  We construct a new statistical measure to determine for the
set of fiducial models the correct value of the parameters for the
transit.  We also show that our algorithm passes the null test,
i.e. it correctly identifies a light-curve with no transit. The set of
fiducial models can be pre-computed and we provide a recipe to do
this. We show that this needs to be done only once before the search
for transits is performed in a set of light-curves.

The speed up in the analysis is significant. For a simulated light-curve
typical of Pan-STARRS we find that our algorithm is $10^8$ times
faster than a search in the full $\chi2$ space. The speed up is due to
the fact that using MOPED the maximum likelihood search is performed on four data (the
number of parameters) instead of thousands and that the ensemble of
fiducial models can be pre-computed.  This achieved increase in speed
to compute the likelihood is important for transit analysis since the
likelihood surface has {\em multiple} maxima, of which only one is the
desired solution and therefore the search for this best solution needs
to explore the whole likelihood surface \footnote{In surveys with 
high cadence and short observational period (e.g. TrES) the likelihood 
surface is smooth and methods utilizing smart searches of the likelihood
surface are better suited.}

This paper is organized as follows: In Section 2, we briefly describe MOPED. 
Section 3 presents the transit model used and how a set of synthetic light-curves 
were constructed. In Section 4, we describe the extension of MOPED using an ensemble of fiducial 
models and  we also present how the results should be compared to the null hypothesis. 
Results are discussed in Section 5  and our conclusions summarized in Section 7. In Section 6, we
describe the numerical topics including a numerical recipe.

\section{MOPED}

We briefly review the parameter estimation and data compression
method MOPED which is originally described in
\citet{HJL00}. The method is as follows: given a set of data {\bf x}
(in our case a light-curve) which includes a signal part ${\bbmu}$ and
noise ${\bf n}$, i.e. 
\be
\bx = \bbmu + \bn,
\ee the idea then is to find
weighting vectors ${\bf b}_m$ where $m$ runs from 1 to the number of
parameters $M$, such that 
\begin{equation} 
	y_m \equiv \bf{b}_m \bf{x}
\end{equation}
 contain as much information as possible about the parameters
(period, duration of the transit etc.).  These {\it numbers} $y_m$ are
then used as the data set in a likelihood analysis with the consequent
increase in speed at finding the best solution. In MOPED, there is one
vector associated with each parameter.

In \citet{HJL00} an optimal and lossless method was found to calculate ${\bf
b}_m$ for multiple parameters (as is the case with transits).
The definition of lossless here is that the Fisher matrix at the
maximum likelihood point is the same whether we use the full dataset
or the compressed version. The Fisher matrix is defined by:
\begin{equation}
\bF_{\alpha\beta} \equiv - \left\langle {\partial^2 \ln {\cal L}\over
\partial \theta_\alpha \partial \theta_\beta} \right\rangle
\end{equation}
where the average is over an ensemble with the same parameters
($\theta_\alpha; \theta_\beta$) but different noise. The a posteriori
probability for the parameters is the likelihood, which for Gaussian
noise is
\begin{eqnarray}
{\cal L}(\theta_\alpha) &=& {1\over (2\pi)^{N/2}
\sqrt{\det(\bC)}}
\times \nn & & 
\exp \left[-{1\over 2}\sum_{i,j} (x_i -
\mu_i)\bC^{-1}_{ij} (x_j-\mu_j)\right].
\end{eqnarray}

The Fisher matrix gives a good estimate of
the errors on the parameters, provided the likelihood surface is well
described by a multivariate Gaussian near the peak.  The method is
strictly lossless in this sense provided that the noise is independent
of the parameters, and provided our initial guess of the parameters is
correct.  This is not exactly true because our initial guess is
inevitably wrong.  However, the increase in parameter errors is very
small in these cases (see \citet{HJL00}) - MOPED recovers the correct solutions
extremely accurately even when the conditions for losslessness are not
satisfied.  The weights required are
\begin{equation}
\bb_1 = {\bC^{-1} \bbmu_{,1}\over \sqrt{\bbmu_{,1}^t
\bC^{-1}\bbmu_{,1}}}
\label{Evector1}
\end{equation}
and
\begin{equation}
\bb_m = {\bC^{-1}\bbmu_{,m} - \sum_{q=1}^{m-1}(\bbmu_{,m}^t
\bb_q)\bb_q \over
\sqrt{\bbmu_{,m}^t \bC^{-1} \bbmu_{,m} - \sum_{q=1}^{m-1}
(\bbmu_{,m}^t \bb_q)^2}} \qquad (m>1).
\label{eq:bbm}
\end{equation}
where a comma denotes the partial derivative with respect to the
parameter $m$ and $\bm{C}$ is the covariance matrix with components
$\bm{C}_{ij}=\langle n_in_j\rangle$.  $i$ and $j$ runs from 1 to the
size of the dataset.  To compute the weight vectors requires an
initial guess of the parameters.  We term this the
fiducial model ($\vec{q}_f$) and we discuss in \S~\ref{sec:performance} the impact
on the MOPED solution on the choice of the fiducial model. For the
case of transits the {\bf C} does not depend on the parameters and
therefore the {\bf b}$_m$ depend only on the fiducial parameters
($\vec{q}_f$).  On the other hand the $\bm{\mu}$ represents the signal part
and thus depends on the free parameters, which we denote by $\vec{q}$.

The dataset $\{y_m\}$ is orthonormal: i.e. the $y_m$ are
uncorrelated, and of unit variance. $y_m$ have means 
\be \langle y_m \rangle = \bb_m( \vec{q}_f) \cdot
\bbmu (\vec{q})
\ee
 The new likelihood is easy to
compute, namely, 
 \begin{eqnarray}
 \ln{\cal L}(\theta_\alpha) &=& {\rm constant} -
\sum_{m=1}^{M} {(y_m-\langle y_m\rangle)^2\over 2} \nonumber \\ 
&=& {\rm constant} -	\sum_m \left[  {\bf b}_m(\vec{q}_f) \cdot {\bf x} - {\bf b}_m(\vec{q}_f) \cdot \bm{\mu}(\vec{q} ) \right]^2 \nonumber \\
\label{eq:MOPEDLike}
\end{eqnarray} 
\begin{figure}
\bc
\includegraphics[width=3.5in]{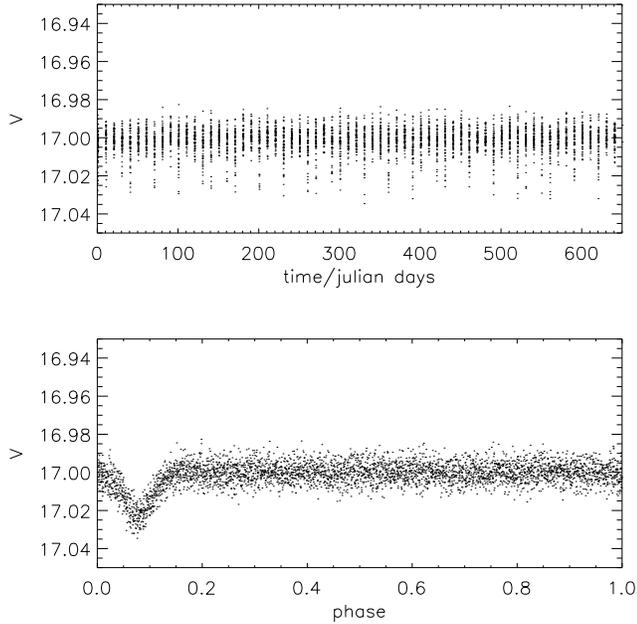}
\caption{Top panel:
Synthetic light curve with transit signal; S/N $\sim 5$,
period $T=1.3$ days, transit duration of $\eta=0.1$ days, transit
depth $\theta \sim 0.01$. Bottom panel: Same synthetic light curve
folded with the right period $T=1.3$ days. The folded light-curve 
has not been used in any of the analysis and it is only shown here for demonstration.
\label{fig:syn}}
\ec
\end{figure}

Further
details are given in \citet{HJL00}.

It is important to note that if the covariance matrix is known for a
large dataset (e.g. a large synoptic
survey) or it does not change significantly from light-curve to light-curve, 
then the $\langle y_m \rangle$ need be computed only {\em once}
for the whole dataset, thus massively speeding up the computing of 
the likelihood. 

\section{Transit model and synthetic light-curves}

\subsection{Transit model}

For the transit analysis we have constructed a model ,$\bm{\mu}$, that
closely represents the shape of a planetary transit light-curve. An
obvious and usually chosen approach is to use a square wave:
$\mu(t)=-1 \,\,-1<t<1$ and $\mu(t)=0$ otherwise. However in order to
allow for softer edges and being analytically differentiable we used
the following function:
\begin{eqnarray}
\mu(t; T, \eta, \theta, \tau)&=& {\rm constant} +\frac{1}{2} \theta \left\{
     2 -{\rm tanh}\left[ c\,(t'+\frac{1}{2}) \right] \right. \nonumber \\
      &+& \left. {\rm tanh}\left[ c\,(t'-\frac{1}{2}) \right]
    \right\} \,\,\,,
    \label{eq:trfid}
\end{eqnarray}
where $t'$
\begin{equation}
    t'(T,\eta, \tau) = \frac{T \, {\rm sin}\left(
        \frac{\pi \, ( t - \tau) }{T} \right) }{\pi \eta} \,\,,
\end{equation}
$T$ is the period, $\tau$ is the epoch, $\eta$ is the transit
duration $\theta$ is the depth of the transit and $c$ is a constant\footnote{
$c$ controls the sharpness of the edges. We used $c=100$ for all calculations
in this work}.


\begin{figure*}
\bc
\includegraphics[width=18cm]{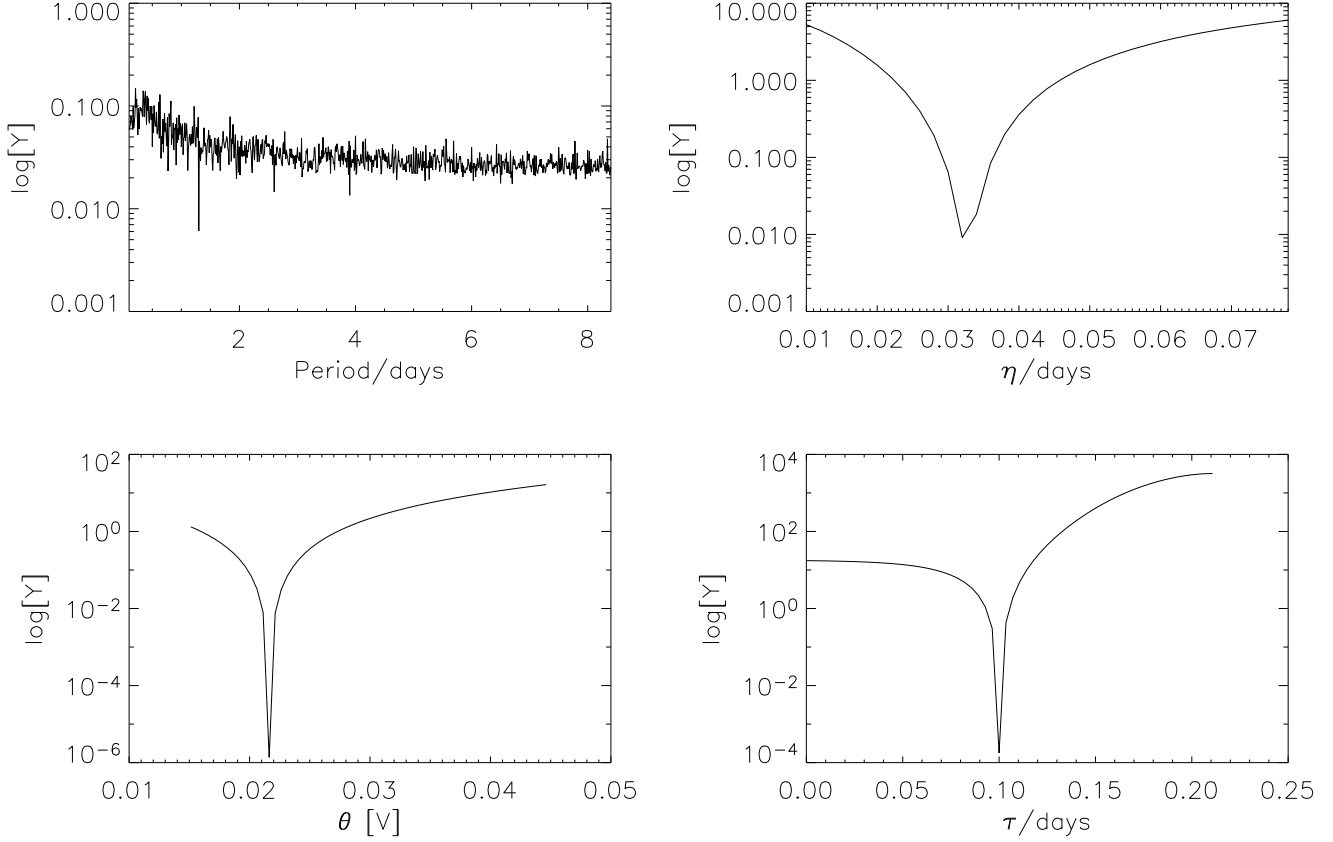}
\caption{ Top panel left:
Likelihood as a function of period $T$. Top panel right:
Likelihood as a function of transit duration $\eta$. Bottom panel
left: Likelihood as a function of $\theta$ and bottom panel right :
Likelihood as a function of $\tau$. In all parameters the correct value is found. 
Note that for $T$ the topology of the Likelihood 
surface is fairly complicated with many local minima, 
thus making efficient minimization techniques not applicable.
\label{fig:transit}}
\ec
\end{figure*}

Applying the transit model to the MOPED framework, one needs to calculate the $\bb$
weight vectors (eq.~\ref{eq:bbm}), which depend
on the derivatives of the model $\bbmu$ 
(the derivatives of eq.~\ref{eq:trfid} with respect to four parameters T, $\theta$, $\eta$,
$\tau$). These derivatives can be analytically calculated and thus
computationally inexpensive since they do not require conditional
statements.

\subsection{Synthetic light-curves}
\label{sec:synthetic}
In order to test our method and estimate the gain in speed we
created a sample of synthetic light-curves by setting the four free
parameters to realistic values and generating magnitudes according to
eq.~\ref{eq:trfid} with Gaussian  noise added to better simulate real light-curves. 
We adjusted the Gaussian noise to achieve desirable signal-to-noise (S/N) values.

We simulated observational sampling patterns 
from Pan-STARRS 
(one observation every 10 minutes, four times a month)
and generated magnitudes as described in
the following equation
\begin{equation}
    x(t_{i}; T, \eta, \theta, \tau) =  \mu(t_{i}; T, \eta, \theta,
    \tau) + \bn_{i}
\end{equation}
where $t_{i}$ are the observational times and $\bn_{i}$ is a Gaussian noise
obtained from Pan-STARRS photometric accuracy of 0.01 magnitudes. 
Fig.~\ref{fig:syn} top panel shows a typical synthetic light-curve 
with period 1.3 days and S/N=5.

\section{Extension to MOPED using an ensemble of fiducial models}
\label{sec:confidence}
Unlike the case of galaxy spectra \citep{HJL00}, the fiducial model will
weigh some data high, very erroneously if the fiducial model is way off from
the true model. This is because the derivatives of the fiducial model with respect 
to the parameters are large near the walls  of the box-like shape 
of the model. 

In this section we present an alternative approach to find the best
fitting transit model to a light-curve.  The method is based on using
an ensemble of randomly chosen fiducial models.  For an arbitrary
fiducial model the likelihood function (eq.~\ref{eq:MOPEDLike}) will
have several maxima one of which is guaranteed to be the correct
solution. This is the case where the values of the free parameters
($\vec{q}$) are close to the true one; thus $\bf{\mu}(\vec{q}$) in
eq.~\ref{eq:MOPEDLike} is similar to $\bf{x}$.  For a different
arbitrary fiducial model there are also several maxima, but only one
will be guaranteed to be a maximum, the true one.  Therefore by using
several fiducial models one can eliminate the spurious maxima and
keep the one that is common to all the fiducial models which is the
true one.  We combine the MOPED likelihoods for different fiducial
models by simply averaging them\footnote{This is 
chosen ad hoc.  We have tried other approaches all of which work
similarly well.  Averaging turned out to be the functional form in
which, error and confidence level of the measurement, could be
easily and analytically calculated.}

\begin{figure}
\bc
\includegraphics[width=8.5cm]{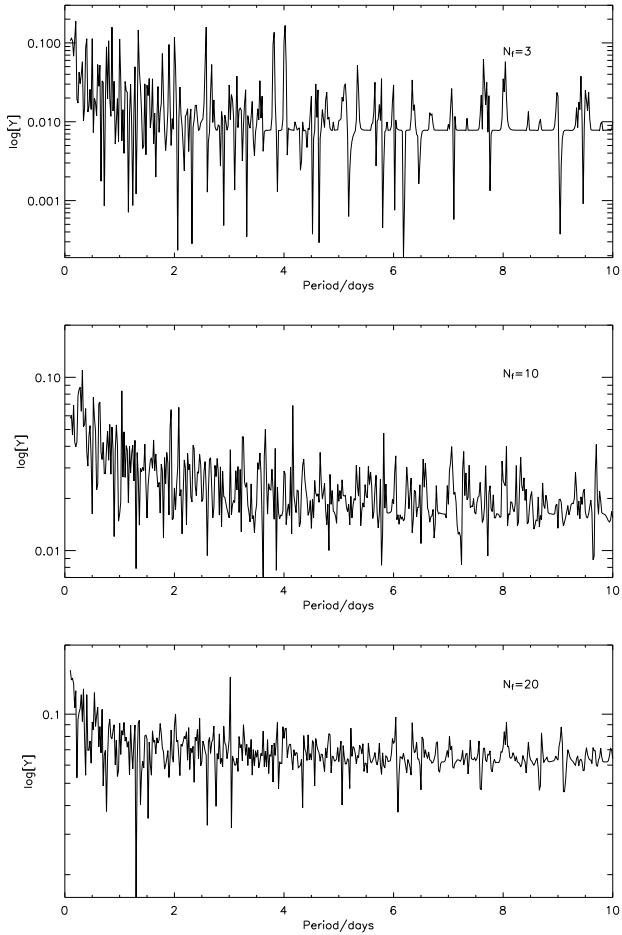}
\caption{$Y$ as a function of period $T$ for
a set of fiducial models for a synthetic light-curve with S/N=3 and 2000 observations
and $T=1.3 \rm{days}$.
The top panel shows the value of $Y$ using
3 randomly selected fiducial models, the middle panel 10 and the bottom using 20. As the number
of fiducial models used increases the number of minima decreases. At $N_f=20$ there is only
one obvious minima at $T=1.3$ days. }
\label{fig:ypmax}
\ec
\end{figure}

The new measure $Y$ is defined:
\begin{equation}
 Y(\vec{q}) \equiv \frac{1}{N_f}\sum_{ \{\vec{q}_f\}} {\cal L}(\vec{q}; \vec{q}_f) \,\,\, ,
\label{eq:y}
\end{equation}
where $\vec{q}$ and $\vec{q}_f$ are the parameter vectors $\{T,\eta,
\theta,\tau\}$ and their fiducial values $\{T_f,\eta_f,
\theta_f,\tau_f \}$ and $N_f$ is the number of fiducial models. 
The summation is over an ensemble of fiducial
models $\{\vec{q}_f\}$.   ${\cal L}(\vec{q};\vec{q}_f)$ is the
MOPED likelihood (eq.~\ref{eq:MOPEDLike}), i.e.

\begin{equation}
   {\cal L}(\vec{q};\vec{q}_f)  =\sum_m \left[  \bm{b}_m(\vec{q}_f) \cdot \bm{x} -  \bm{b}_m(\vec{q}_{f}) \cdot \bm{\mu}(\vec{q}) \right]^2
\label{eq:chi2}
\end{equation}
Fig.~\ref{fig:ypmax} shows the $Y$ as a function of period $T$ for 
a different size sets of fiducial models for a synthetic light-curve with S/N=3 and 2000 observations. 
The top panel shows the value of $Y$ using 
an ensemble of 3 fiducial models. As it can be seen from the figure there are more
than few minima. Using an ensemble of 10 fiducial models (shown in the next panel)  
reduces the number of  minima. In the last panel we used an ensemble of 20 fiducial
models and there is only one obvious minimum, the true one. 

Fig.~\ref{fig:transit} 
shows the value of $Y$ as a function of each free parameter
for a synthetic light-curve. 
We set the values of 3 of the parameters to the ``correct'' values
(used to construct the light-curve) and we let the fourth free for each panel. Note that the shape of 
the $Y$ as a function of $\eta$, $\theta$ and $\tau$ is 
smooth, however the dependency on $T$ is erratic suggesting that
efficient minimization techniques are not applicable. 

\subsection{Confidence and error analysis}
\label{sec:conf}
To confidently determine that the minimum found is not spurious 
the likelihood of the candidate solution must be compared to the value and distribution
of $Y$ derived from a set of light-curves with no transit signal.
One can simulate a set of null light-curves and build a distribution by calculating  the 
value of $Y$  for each point in the parameter space for each simulated ``null'' light-curve; 
a real expensive computational task. Alternatively this null distribution can be analytically derived. 

Since  $x\sim N(\langle x \rangle, \sigma_x)$ and all other variables are deterministic,
then it can be shown that $Y(\vec{q})$ follows a non-central ${\cal X}^2$
distribution $Y(\vec{q}) \sim {\cal X}^2(r,\lambda)$ where
$r$  is the number of degrees of freedom and $\lambda$ is the non centrality
of the distribution. The non-central ${\cal X}^2$ distribution has 
mean and variance according to: 
\begin{eqnarray}
  \mu  &=& r+\lambda \,\,\,,\\
  \sigma^2 &=& 2(r+2\lambda) \,\,\, ,
\end{eqnarray}
 where $r=4$ and  $\lambda$ is given by  
\begin{equation}
  \lambda = \frac{{\rm E}^2 \left[ {\cal X} \right]}{{\rm var}\left[ {\cal X} \right]} \,\,.
\end{equation}
%
%
%
%
The square of the expectation value is, 
\begin{equation} 
  \rm{E}^2 \left[ {\cal X} \right] = \sum_m
          \left[ \langle x \rangle \,\, B_m(\vec{q}_f) - C_m(\vec{q};\vec{q}_{f})  \right]^2
\end{equation}
%
%
%
%
where we define 
\begin{equation}
   B_m(\vec{q}_f) \equiv \sum_t b_m^t(\vec{q}_f) ,
\label{eq:defB}
\end{equation}
\begin{equation}
   D_m(\vec{q};\vec{q}_{f}) \equiv  \bm{b}_m(\vec{q}_{f}) \,\cdot  \bm{\mu}(\vec{q}) 
\end{equation}
%
%
%
%
%
and  the variance is given by
\begin{eqnarray}
  {\rm var} \left[ {\cal X} \right]  &=&  {\rm var} \left[ \sum_{m} \bm{b}_m(\vec{q}_f) \cdot \bm{x} - \sum_{m} \bm{b}_m(\vec{q}_{f}) \cdot \bm{\mu}(\vec{q}) \right] \nonumber \\
   &=&  \sum_{m} \left| \bm{b}_m(\vec{q}_f) \right|^2 {\rm var}\left[ x^t \right] \nonumber \\
   &=& \sigma^2_x \beta_m(\vec{q}_f)
\end{eqnarray}
where we define $\beta_m(\vec{q}_{f})$ to be:
\begin{equation}
   \beta_m(\vec{q}_{f}) \equiv   \bm{b}_m(\vec{q}_{f}) \cdot \bm{b}_m(\vec{q}_{f})  \,\,\,.
\end{equation}
%
%
%
Combining the above equations we get 
\begin{equation}
  \lambda = \frac{ \sum_m\left[ \langle x \rangle \,\, B_m(\vec{q}_f) - D_m(\vec{q};\vec{q}_{f})  \right]^2}{ \sigma^2_x \, \beta_m(\vec{q}_f)}
\label{eq:lambda}
\end{equation}
\noindent
To compute confidence levels for a particular $Y$ 
we integrate a non-central ${\cal X}^2$ distribution
with non centrality given by eq.~\ref{eq:lambda} from $Y(\vec{q})$ to
infinity. This is done numerically, still this is a very quick 
operation. Furthermore, as we will show in sec.~\ref{sec:performance} 
this will only be performed few times per light curve.  

\begin{figure}

\includegraphics[width=8.5cm]{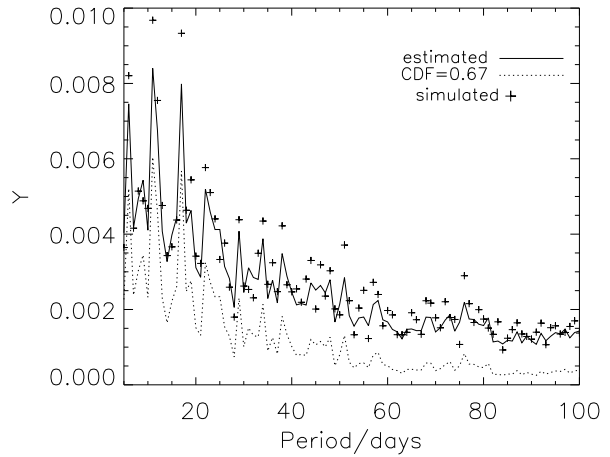}
\caption{ Values of $Y(T)$ for
the null case (i.e. a light-curve without a transit) both simulated
(crosses) and analytically calculated (see sec.\ref{sec:conf})  
(solid line is the expected value and dotted line is 
the $67\%$ confidence level). It is clear that the simulated values
agree well with the theoretical ones.  
\label{fig:compy}
}
\end{figure}

Fig.~\ref{fig:compy} shows the values of $Y(T)$ for
the null case (i.e. a light-curve without a transit) both simulated
(crosses) and theoretically calculated using the equations above 
(solid line is the expected value and dotted line is 
the $80\%$ confidence level). It is clear that the simulated values
agree well with the theoretical ones. 
Note that because the confidence can be calculated analytically we 
do not have to simulate null light-curves and recalculate the $Y$ 
for each light-curve thus gaining computational speed.

\begin{figure}
\includegraphics[width=8cm]{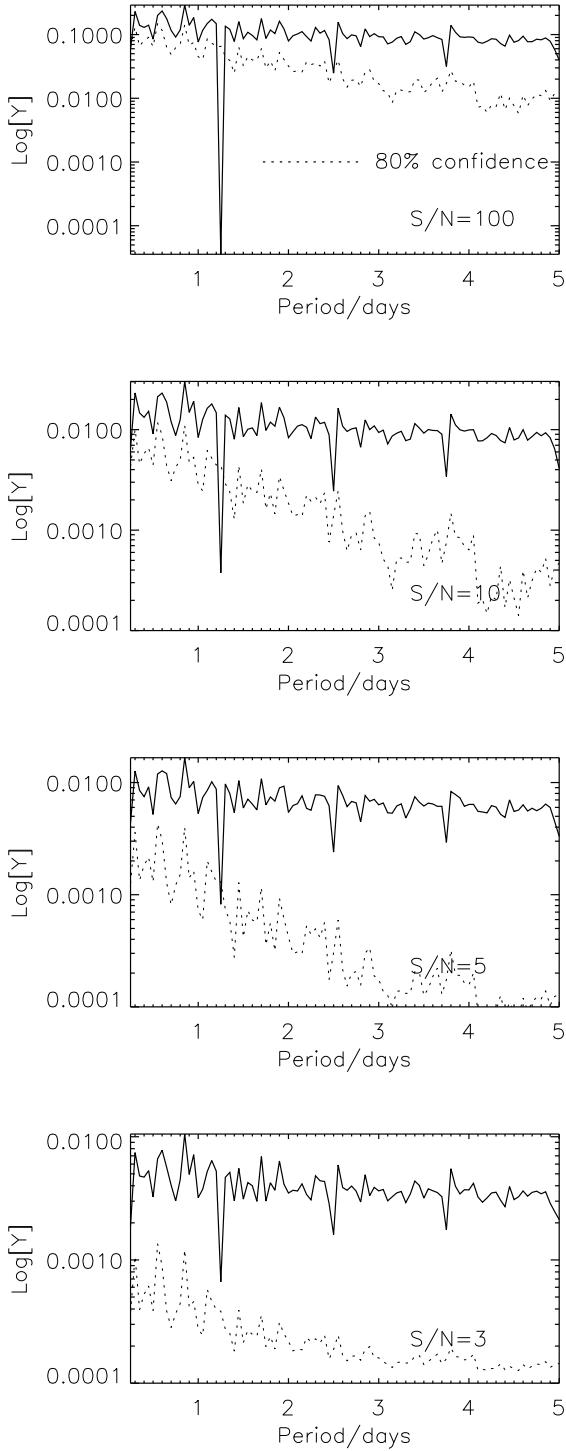}
\caption{The value of $Y$ is shown as a function of period for a
synthetic light-curve with a transit at $1.25$ days. The different panels
show different values of $S/N$. Note that there is a well defined
minimum at the right period. The dotted line shows the 80\% confidence
level. Note that at this level there is only one single minimum at the
right period even for $S/N$ as low as 5.
\label{fig:simvsest}
}
\end{figure}

\section{Results}

Fig.~\ref{fig:transit} shows the results of likelihood as a function
of each parameter using a typical synthetic light-curve. 
The above searches were performed only in one parameter at the time,
irregardless we successfully recover the true values for the parameters of the transit. 

In Fig.~\ref{fig:simvsest} we show the value of $Y$ as a function of period for a
synthetic light-curves with a transit of 1.28 days. The run was done using 
40 fiducial models. The different panels
show different values of $S/N$. The dotted line shows the 80\% confidence
level. For all 4 cases there is a well defined
minimum at the right period, where the minimum is below the $80\%$ level 
for $S/N$ as low as 5 and at 71$\%$ for  $S/N=3$.

\begin{figure}
\includegraphics[width=9.0cm]{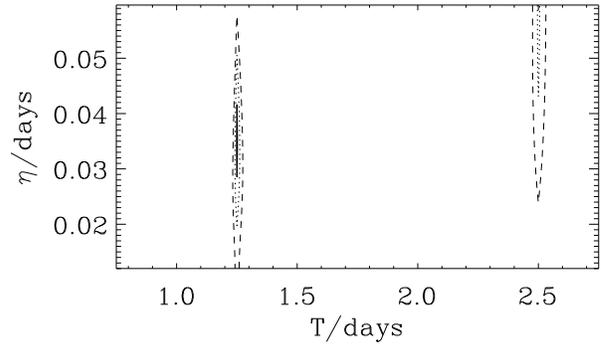}
\caption{Projection of the four-dimensional likelihood surface on the
two dimensional space $\tau-$T, note that contours close around the
right period and that they appear at multiples of the right period as
happens for the one-dimensional case (see text for more details).}
\label{fig:theta_tau}
\end{figure}

\begin{figure}
\includegraphics[width=9.0cm]{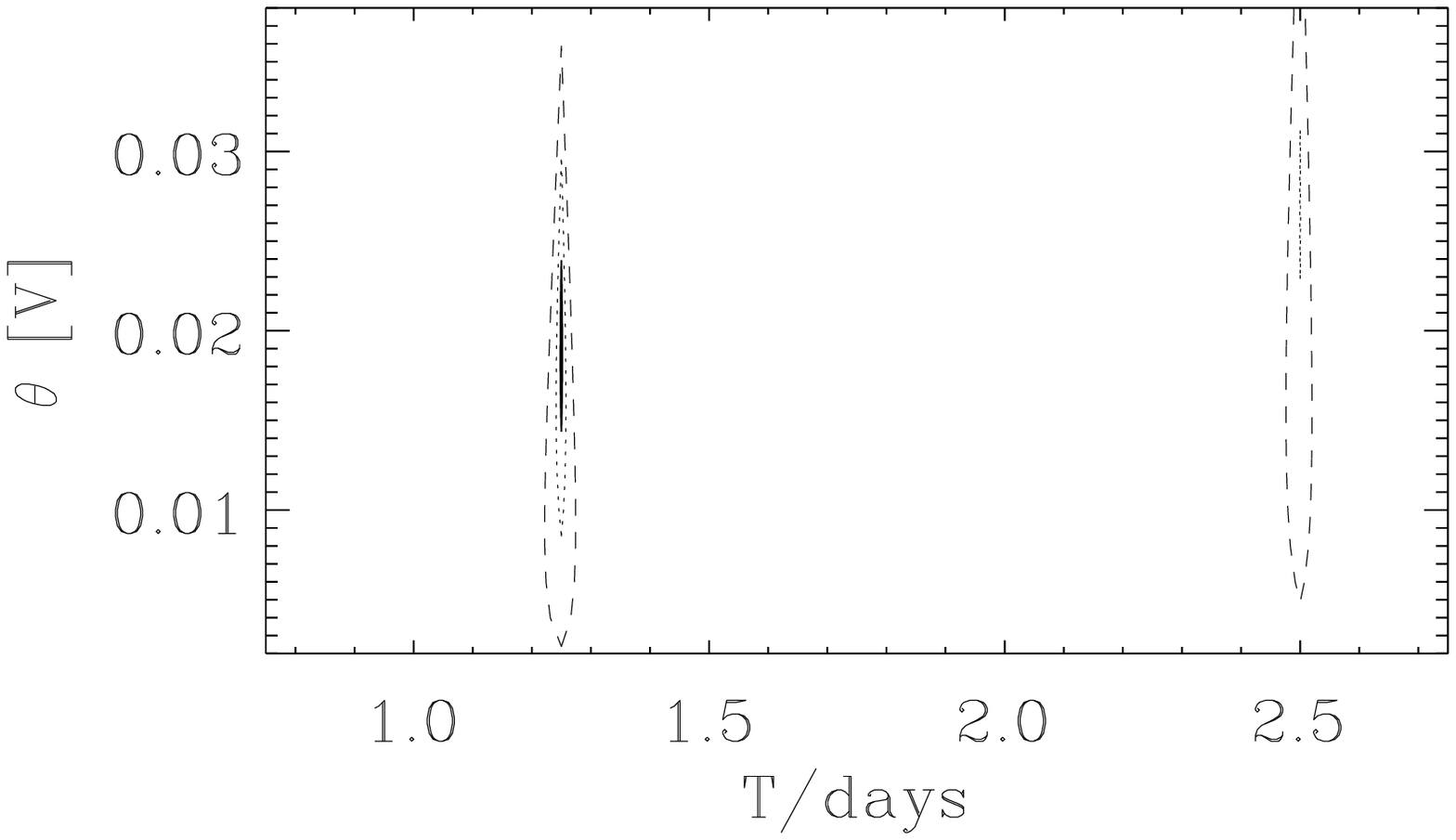}
\caption{Same as fig.~\ref{fig:theta_tau} but for parameters $\theta$ and T.}
\label{fig:eta_theta}
\end{figure}

\begin{figure}
\includegraphics[width=9.0cm]{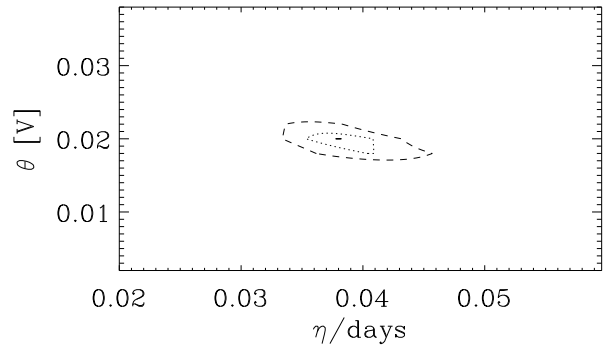}
\caption{Same as fig.~\ref{fig:theta_tau} but for parameters $\theta$ and $\tau$}
\label{fig:eta_tau}
\end{figure}

The more realistic case is to perform the search in the four parameter
space simultaneously and show that our method successfully recovers
the ``correct" values of $T$, $\eta$, $\theta$ and $\tau$ for a sample
of synthetic light curves. This is shown in figures
\ref{fig:theta_tau}, \ref{fig:eta_theta} and \ref{fig:eta_tau} where
the 2D projections of the four dimensional search are presented. The
different contours correspond to $50\%$, $65\%$ and $80\%$ confidence
levels. It is worth commenting the ``multiple'' maxima in the
likelihood. This feature also appears in the one-dimensional search:
multiple minima appear at multiples of the true period, but note that
the best fitting model is still the true period only (at the $50$\%
confidence level the other solutions are excluded). This behavior is
expected since when the period is allowed to be a multiple of the true
one, one out of $n$ ($n$ is an integer) transits will fit and
therefore will produce a better fit than the null case. These multiple
solutions can be easily excluded by keeping the shortest period. This 
only occurs for $T$, the other parameters have only one well
defined minimum at the true value.


\subsection{Application to multiple light-curves}
\label{sec:mult}

After having shown that our algorithm works properly on several
synthetic light-curves, we now explore the performance of the method
for a wide range of values for T, $\theta$, $\eta$ and $\tau$. In
particular we have simulated light-curves for $0.1 < T < 4$ days; $1\%
< \eta < 5\%$ of the period; $0 < \tau < T$ and
$R_{\rm{planet}}/R_{\star} \sim 0.1$ and $12 < V < 24$. The
observation frequency of the light-curve is similar to that of a
Pan-STARRS light curve. This space parameter and observation frequency
should cover the range of light transit observations expected from
surveys like Pan-STARRS ({\tt http://pan-starrs.ifa.hawaii.edu}) and
LSST ({\tt http://www.lsst.org}).

\begin{figure}
\includegraphics[width=8.5cm]{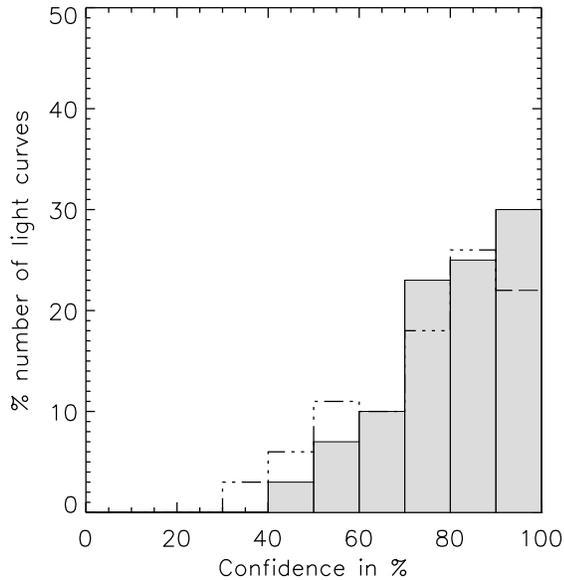}
\caption{Distribution function for the confidence of finding the true
period $T$ for a set of $100$ simulated light curves with $S/N=5$ for
the range of values of transit parameters described in
\S~\ref{sec:mult}). Two sets of $100$ light curves have been
simulated: a set were the number of measurements per light curve is
$2000$ (dotted line) and another with $4000$ measurements (solid
line). Note that for the $4000$ measurements case most $T$s are
recovered with confidence higher than $80$\% and that for about $50$\%
of the simulated light curves the confidence in recovering the true
period is greater than $95$\%.}

\label{fig:hist_conf}
\end{figure}

We have simulated 100 light-curves with $S/N=5$.  For each light-curve
we estimated the likelihood $Y(\vec{q}_{\rm{true}})$ for the
ensemble of fiducial models and then we calculated the confidence that
this value is not a spurious detection. Fig.~\ref{fig:hist_conf}
shows the distribution for these confidence values.  Two curves are
plotted: the dotted line is for curves with a total number of $2000$
observations (about 1 year).  The thick solid line is for the case
when the range is doubled to 4000 measurements (2 years). For the
higher number of observations there is a significant increase in the
confidence of recovering the true period. For this case most transits
($80\%$) are found with confidence over the null case higher than $70$\%,
i.e. for all galaxies the recovered period $T$ has a confidence
greater than $70$\% of being the correct one. In about $25$\% of the
cases the confidence in recovering the true period is greater than
$90$\%. For the case of $2000$ measurements the success rate is
somewhat lower.
This is because the error on the estimated parameters depends on the number of 
observations. For $\theta, \eta$ and $\tau$ 
this depends on the number of observations in the transits. 
However, for the $T$ this depends on the number of transits observed.
One can show that the probability of
observing a single transit is proportional to $\eta/T$, thought the
probability of observing multiple transits is smaller. Furthermore, it
also depends on the irregularity of the observational times (the more
irregular the times the better the chance of recovering the
signal).

\section{Numerical Method}
\label{sec:performance}
The real advantage of the present method lies on the fact that for a
set of light-curves most of the work can be done once in calculating
the fiducial models. In this section we describe in more detail the
numerical approach and we present the numerical gain over the brute
force calculation using the full ${\cal X}^2$.

\subsection{Calculations of the fiducial models}
The second term of equation \ref{eq:chi2} does not depend on the
actual light-curve data $\bm{x}$. Therefore,
$D_m(\vec{q}_{f};\vec{q})$, $b_m^t(\vec{q}_f)$ and
$\beta_m^t(\vec{q}_f)$ for each fiducial model can be pre-calculated
once and stored in files.  Thus for each light-curve we only need to
calculate $\sum_t b_m^{t}(\vec{q}_f) \, x^t$, which is independent of
the search parameter $\vec{q}$. This is a major advantage of our
method. Before we describe the numerical steps in more detail we need
to address how we choose the fiducial models and how many fiducial
models are needed.

\subsection{Choice of fiducial models}
There are three questions that we need to address about the choice of 
fiducial models. 

\underline{Number of fiducial models:} Since the confidence level can be 
calculated at each iteration step, the number of fiducial models
do not need to be pre-determined. If there is only one solution at confidence
larger than 70$\%$ that parameter is considered to be  the correct value 
and the iteration is stopped.  Yet, for light-curves with
low S/N the actual solution may never exceed that threshold. Therefore
we imposed a maximum of  100 fiducial models. Besides, for a
typical survey there will be only few expected light-curves containing
a transit signal, thus for most cases the iteration will be terminated 
at the 100 fiducial model limit.

\underline{Choice of fiducial parameters:} 
We have found that the best performance in finding the true solution 
comes when we choose fiducial 
models from a flat distribution spanning the full range 
of the free parameters.

\underline{Choice of search parameters $\vec{q}$:} 
Despite the fact that $D_m(\vec{q}_{f};\vec{q})$  will be calculated only once 
and it will not contribute to the overall computational burden 
of finding transits in a set of light-curves, the size of the database (files)
that stores the fiducial model information heavily depends on the choice 
of the free parameters range and grid size. This is mostly important 
for future space missions where the memory available is limited. 

As it can be seen from Fig.~\ref{fig:transit} (top-left panel)
finding the ``correct" period where there is a non linear relationship
between likelihood and period is the most difficult task.  This is due
to the fact that a small change in the value of period $T$ produces a
huge variation at the tail of the light-curve.

Theory suggests that the asymptotic standard deviation of the estimate
of the period is of the order $T^{-2/3}$, so the grid should be that
small too.  We therefore performed the search on a uniform grid in
frequency, $T^{-1}$, rather than on a uniform grid in $T$.  There is
also a related question of how fine the search grid should be for
$\eta$ and $\tau$. Since for data folded at period $T$, the folded
observation times are roughly uniform, the average spacing of
subsequent folded observations is $T/N$ ($N$ is the total number of
observations), $T/N$ is a natural choice for the grid size for the
$\eta$ and $\tau$ searches.  For a typical search the total number of
searches can be as high as $10^9$ which translates to $1$TB of
data. This is prohibiting for space missions. In what follows we
examine how to further reduced the search space using physical and
statistical arguments.

{\em Transit length range:} For a given period one can allow $\eta$
to take values between 0 and $T/2$. This is a naive estimate based on
the fact that the planet spends half of the time in front of the star.
The range of $\eta$ can be further limited using geometrical arguments
and Kepler's law. It can be shown that the transit duration is
\citep{Sackett1999}
\begin{equation}
    \eta \approx \frac{T}{\pi} \sqrt{ \left( \frac{\rstar}{a}
    \right)^2 - cos^2(i) }
\end{equation}
where $\rstar$ is the radius of the star, $a$ is the orbit radius
of the planet and $i$ is the inclination angle. The maximum value
that $\eta$ can take is when the inclination angle $i$ is zero.
Using Kepler's law the ratio of duration over period is
\begin{equation}
    \frac{\eta}{T} = \frac{R}{\pi\,\left(
    \frac{T^{2}GM_{\star}}{4\pi^2} \right)^{1/3}}
    \label{eq:duroverperiod}
\end{equation}
For a typical main sequence star this yields a $\eta/T \approx
4\%$ for periods of 1-2 days. The fraction gets smaller as the
period increases resulting a gain of factor of 50 in computational
time (compared to the naive approximation $\eta/T \leq 1/2$).

{\em Longest period}: Equation~\ref{eq:duroverperiod}
can be used to determine the longest period that can be recovered
from the data. Namely at which period the transit duration over the period is small enough
that the probability
of observing more than few occultations is insignificant.  It can be
shown that for most inclinations the probability of observing an
occultation is given by
\begin{equation}
    p = \frac{\eta}{T}=\frac{1}{2\pi}  \frac{\rstar}{a} \,\,.
    \label{eqpt}
\end{equation}
This is basically the probability that the occultation time
and the observation time overlap.

The probability  of observing $x$ occultations during the whole
lifetime of the survey is given by a binomial distribution. At
the limit where the number of observations is large, the
probability distribution becomes a Gaussian distribution

\begin{equation}
    P_{t}(x) = \frac{1}{\sqrt{2\pi \, \sigma^2}}
        \exp^{ -\frac{ (x - \mu)^2 }  { 2 \, \sigma^2}} \,\, ,
\end{equation}
where $p$ is given in eq.~\ref{eqpt}. The mean value is given by
\begin{equation}
    \mu=n_{t}\, p \,\, ,
\end{equation}
and the standard deviation
\begin{equation}
\sigma = \sqrt{n_{t} \, p\, (1-p)} \,\,.
\end{equation}
where $n_{t}$ is the number of complete transits $T_{\rm range}/T$.
The probability of observing at least 3 transits is therefore
given by the integral
\begin{equation}
    P(x\geq3) = \frac{1}{\sqrt{2\pi \, \sigma^{2}}}
       \int_{3}^{\infty} \exp^{ -\frac{ (x - \mu)^2 }  { 2 \,
       \sigma^2}}\,dx \,\,.
\end{equation}
For a typical main sequence star a planet with a period of 20 days
has a probability of observing 10 occultations that is less than
few percent.  Using that as the upper limit to our search reduces
the number of iterations by a factor 5-10.

\subsection{Numerical recipe}
The steps of the numerical method are described below
\begin{enumerate}
	\item Select a set of fiducial models. The 
	choice of the fiducial parameters span the domain
	of the search parameters. 
	\item Calculate   $D_m(\vec{q}_{f};\vec{q})$, $b_m^t(\vec{q}_f)$ and 
$\beta_m^t(\vec{q}_f)$ for each fiducial model. The range and sampling frequency of the free 
parameters $\vec{q}$ are according to the physical arguments described above.   Save values in a database (binary files)
        \item  For each light-curve calculate  $\sum_t b_m^{t}(\vec{q}_f) \, x^t$. 
	\item Search through the fiducial models for $D_m(\vec{q}_{f};\vec{q})$ with 
	similar values as  $\sum_t b_m^{t}(\vec{q}_f) \, x^t$ from previous step. 
	Note that since the database is sorted with respect to $D$,
	this is a $log(N_q)$ operation where $N_q$ is the number of free parameter values. 
      \item Calculate $Y$ for those parameters such that $ \sum_t b_m^{t}(\vec{q}_f) \, x^t - D_m(\vec{q}_{f};\vec{q}) $ is small.
	\item Compute the confidence level for the selected $\vec{q}$'s using
	eq.~\ref{eq:lambda}. Note that since $D$'s, $B$'s and $\beta$'s  are pre-calculated we only need 
	to compute $\langle x \rangle$. 
	\item If there is only one minima with confidence level higher than 70$\%$ exit. 
	\item If number of fiducial models is larger than 100, exit.
	\item Back to (iii).
\end{enumerate}

\subsection{Required number of operations}
The brute force minimization for the likelihood function requires 
\begin{equation}
	N^{\rm{brute}}_{\rm{total}} \sim N_{\rm{obs}} \, N_{q}
\end{equation}
The number of operations for our method after the fiducial models are 
computed is 
\begin{equation}
	N^{\rm MOPED+}_{\rm{total}} \sim N_{\rm{obs}} \, N_{\rm{fid}}
\end{equation}
For a typical light-curve with low observing frequency like Pan-STARRS in the 
the four dimensional parameter space $N_{q}$ can easily be $10^{10}$. 
This number is large because of the non-linear dependence of the period to 
the likelihood, thus  $N_{T} \sim 100000$ (see the arguments above).
Contrary $N_{\rm{fid}} \sim 10^2$ which means an  improvement in speed
of a factor of $10^8$. 

\section{Conclusions}
We have presented a new algorithm to fast and efficiently detect
transits in light-curves. Our algorithm does produce a major speed up
factor in light transit searches, of about eight orders of magnitude,
compared to the brute force method using the full $\chi^2$. This
translates in finding a transit on a light-curve with 10$^4$
observations in well under a second on current desktop computers. We
have developed a four parameter model for the transit of an object and
have shown, using synthetic light-curves, that our algorithm is
successful at recovering the true parameters of the transit. We have simulated a set
of light-curves with the sampling rate and
photometric accuracy expected in large synoptic surveys like
Pan-STARRS and shown that for a large range in the values of the
parameters (T, $\eta$, $\theta$, $\tau$) we recover the true
values. For surveys like Pas-STARRS and LSST it should
be possible to detect transits by Jovian planets and planets several
times the size of earth.  Since the expected detection rate of
transits in this large surveys is very low, only one transit out of
thousands light-curves, we believe that our method provides a fast and
efficient algorithm to detect transits for future surveys.

\section{ACKNOWLEDGMENTS} P. Protopapas wishes to thank Rahul Dave for 
valuable discussions. 


\begin{thebibliography}{}

\bibitem[\protect\citeauthoryear{{Aigrain} \& {Favata}}{{Aigrain} \&
  {Favata}}{2002}]{AigrainFavata02}
{Aigrain} S.,  {Favata} F.,  2002, A\&A, 395, 625

\bibitem[\protect\citeauthoryear{Alonso et al.}{2004}]{Alonso2004} 
Alonso R., et al., 2004, ApJ, 613, L153 

\bibitem[\protect\citeauthoryear{{Charbonneau}, {Brown}, {Latham} \&
  {Mayor}}{{Charbonneau} et~al.}{2000}]{Charbonneau2000}
{Charbonneau} D.,  {Brown} T.~M.,  {Latham} D.~W.,    {Mayor} M.,  2000, ApJL,
  529, L45

\bibitem[\protect\citeauthoryear{{Charbonneau}, {Brown}, {Noyes} \&
  {Gilliland}}{{Charbonneau} et~al.}{2002}]{Charbonneau2002}
{Charbonneau} D.,  {Brown} T.~M.,  {Noyes} R.~W.,    {Gilliland} R.~L.,  2002,
  Apj, 568, 377

\bibitem[\protect\citeauthoryear{Defa{\" y}, Deleuil, \& 
Barge}{2001}]{Defay+01} Defa{\" y} C., Deleuil M., Barge P., 
2001, A\&A, 365, 330 

\bibitem[\protect\citeauthoryear{{Doyle}, {Deeg}, {Kozhevnikov}, {Oetiker},
  {Mart`{\'{\i}}n}, {Blue}, {Rottler}, {Stone}, {Ninkov}, {Jenkins},
  {Schneider}, {Dunham}, {Doyle} \& {Paleologou}}{{Doyle}
  et~al.}{2000}]{Doyle+00}
{Doyle} L.~R.,  {Deeg} H.~J.,  {Kozhevnikov} V.~P.,  {Oetiker} B.,
  {Mart{\'{\i}}n} E.~L.,  {Blue} J.~E.,  {Rottler} L.,  {Stone} R.~P.~S.,
  {Ninkov} Z.,  {Jenkins} J.~M.,  {Schneider} J.,  {Dunham} E.~W.,  {Doyle}
  M.~F.,    {Paleologou} E.,  2000, ApJ, 535, 338

\bibitem[\protect\citeauthoryear{{Gilliland}}{{Gilliland}}{2000}]{Gilliland.20%
00}
{Gilliland} e.~a.,  2000, ApJL, 545, L47

\bibitem[\protect\citeauthoryear{{Heavens}, {Jimenez} \& {Lahav}}{{Heavens}
  et~al.}{2000}]{HJL00}
{Heavens} A.,  {Jimenez} R.,    {Lahav} O.,  2000, MNRAS, 317, 965

\bibitem[\protect\citeauthoryear{{Hoekstra}, {Wu} \& {Udalski}}{{Hoekstra}
  et~al.}{2005}]{Hoekstra.2005}
{Hoekstra} H.,  {Wu} Y.,    {Udalski} A.,  2005, astro-ph/0501353

\bibitem[\protect\citeauthoryear{{Jenkins}, {Caldwell} \& {Borucki}}{{Jenkins}
  et~al.}{2002}]{Jenkins+02}
{Jenkins} J.~M.,  {Caldwell} D.~A.,    {Borucki} W.~J.,  2002, ApJ, 564, 495

\bibitem[\protect\citeauthoryear{{Konacki}, {Torres}, {Sasselov} \&
  {Jha}}{{Konacki} et~al.}{2003}]{Konacki2003}
{Konacki} M.,  {Torres} G.,  {Sasselov} D.~D.,    {Jha} S.,  2003, ApJ, 597,
  1076

\bibitem[\protect\citeauthoryear{{Kov{\' a}cs}, {Zucker} \& {Mazeh}}{{Kov{\'
  a}cs} et~al.}{2002}]{Kovacs+02}
{Kov{\' a}cs} G.,  {Zucker} S.,    {Mazeh} T.,  2002, A\&A, 391, 369

\bibitem[\protect\citeauthoryear{{Mall{\' e}n-Ornelas}, {Seager}, {Yee},
  {Minniti}, {Gladders}, {Mall{\' e}n-Fullerton} \& {Brown}}{{Mall{\'
  e}n-Ornelas} et~al.}{2003}]{Mallen.2003}
{Mall{\' e}n-Ornelas} G.,  {Seager} S.,  {Yee} H.~K.~C.,  {Minniti} D.,
  {Gladders} M.~D.,  {Mall{\' e}n-Fullerton} G.~M.,    {Brown} T.~M.,  2003,
  ApJ, 582, 1123

\bibitem[\protect\citeauthoryear{{Mochejska}, {Stanek}, {Sasselov} \&
  {Szentgyorgyi}}{{Mochejska} et~al.}{2002}]{Mochejska.2002}
{Mochejska} B.~J.,  {Stanek} K.~Z.,  {Sasselov} D.~D.,    {Szentgyorgyi} A.~H.,
   2002, AJ, 123, 3460

\bibitem[\protect\citeauthoryear{{Pont}
  et~al.}{2005}]{Pont.2005}
Pont F., et al., 2005, astro-ph/0501611 

\bibitem[\protect\citeauthoryear{Sackett}{1999}]{Sackett1999} 
Sackett P.~D., 1999, poss.conf, 189 

\bibitem[\protect\citeauthoryear{{Street}, {Horne}, {Lister}, {Penny},
  {Tsapras}, {Quirrenbach}, {Safizadeh}, {Mitchell}, {Cooke} \&
  {Cameron}}{{Street} et~al.}{2003}]{Street+03}
{Street} R.~A.,  {Horne} K.,  {Lister} T.~A.,  {Penny} A.~J.,  {Tsapras} Y.,
  {Quirrenbach} A.,  {Safizadeh} N.,  {Mitchell} D.,  {Cooke} J.,    {Cameron}
  A.~C.,  2003, MNRAS, 340, 1287

\bibitem[\protect\citeauthoryear{{Tingley}}{{Tingley}}{2003}]{Tingley2003}
{Tingley} B.,  2003, A\&A, 403, 329

\bibitem[\protect\citeauthoryear{Torres et al.}{2004}]{Torres.2004} 
Torres G., Konacki M., Sasselov D.~D., Jha S., 2004, ApJ, 614, 979 

\bibitem[\protect\citeauthoryear{{Udalski}, {Paczynski}, {Zebrun}, {Szymaski},
  {Kubiak}, {Soszynski}, {Szewczyk}, {Wyrzykowski} \& {Pietrzynski}}{{Udalski}
  et~al.}{2002}]{Udalski.2002}
{Udalski} A.,  {Paczynski} B.,  {Zebrun} K.,  {Szymaski} M.,  {Kubiak} M.,
  {Soszynski} I.,  {Szewczyk} O.,  {Wyrzykowski} L.,    {Pietrzynski} G.,
  2002, Acta Astronomica, 52, 1

\bibitem[\protect\citeauthoryear{{Udalski}, {Pietrzynski}, {Szymanski},
  {Kubiak}, {Zebrun}, {Soszynski}, {Szewczyk} \& {Wyrzykowski}}{{Udalski}
  et~al.}{2003}]{Udalski.2003}
{Udalski} A.,  {Pietrzynski} G.,  {Szymanski} M.,  {Kubiak} M.,  {Zebrun} K.,
  {Soszynski} I.,  {Szewczyk} O.,    {Wyrzykowski} L.,  2003, Acta Astronomica,
  53, 133

\end{thebibliography}

\end{document}